\documentclass[12pt,preprint]{aastex}
\shorttitle{Orientation Coherence and the Substructure Abundance}
\shortauthors{Shim \& Lee}
\begin{document}
\title{Dependence of the Substructure Abundance on the Orientation Coherence of the Halo Tidal Field}
\author{Junsup Shim and Jounghun Lee}
\affil{Astronomy program, Department of Physics and Astronomy,
Seoul National University, Seoul  08826, Republic of Korea \\
\email{jsshim@astro.snu.ac.kr, jounghun@astro.snu.ac.kr}}
\begin{abstract}
A numerical evidence for the dependence of the substructure abundance of cluster halos on the orientation coherence 
of the surrounding tidal fields is presented.  Applying the adapted minimal spanning tree (MST) algorithm to the cluster halos with 
$M\ge 10^{14}\,h^{-1}\,M_{\odot}$ from the Big MultiDark-Planck Simulations, we identify primary MST stems composed of multiple 
nodes as the filaments and measure their specific sizes (spatial extents per node), which quantify the orientation coherence of the 
tidal field on the cluster scales. Classifying the cluster halos into five samples by the specific sizes of their host 
filaments and tallying the mass distributions of the five samples, we investigate if and how the substructure abundance of the cluster 
halos differs among the five samples. It is found that the cluster halos embedded in the filaments with larger specific sizes tend to 
possess less substructures. This anti-correlation is also shown robust against narrowing down the formation epochs of 
the cluster halos as well as against fixing the node number of the filaments. Given that the filaments with larger specific sizes 
form at the regions where the surrounding tidal fields are coherent over large scales in the orientations of their principal axes, 
we suggest that in the filaments with larger specific sizes the satellite infall and matter accretion onto the cluster halos should be 
obstructed due to the development of the tangential velocities in the plane perpendicular to the elongated axes of the filaments. 
\end{abstract}
\section{Introduction}\label{sec:intro}

The substructure of dark matter halos has been the subject of many literatures in which analytical and/or numerical methods were 
employed to investigate such intriguing topics as the mass and spatial distributions of substructures, the dependence of substructure 
abundance on the large-scale density and tidal fields as well as on the host halo mass, the effects of tidal stripping and dynamical 
friction on the survival rates of substructures and so on 
\citep[e.g.,][]{tor-etal98,sheth03,lee04,luc-etal04,NS04,OL04,van-etal05,hah-etal09,HT10,gao-etal11,til-etal11,
wan-etal11,cro-etal12,wu-etal13,XG15,JV17,zomg1,zomg3}.
Owing to those previous endeavors, a coarse roadmap is now established to explain the presence and properties of 
the substructures of dark matter halos in the standard $\Lambda$CDM ($\Lambda$ + Dold Dark Matter) universe.  

The presence of substructures in a bound halo is a natural consequence of hierarchical merging process. 
Once a substructure forms, it is prone to severe mass-loss due to various dynamical effects such as tidal stripping, close 
encounter, and dynamical friction inside its host halo \citep[e.g.,][]{tor-etal98,TB04}.   
The longer a substructure is exposed to those dynamical effects, the harder it can survive.  Thus,  at fixed mass a halo that 
forms earlier tends to contain less number of substructures (or equivalently, less mass fraction) \citep[e.g., see][]{JV17}. 
Among various factors that affect the formation epoch of a halo, the most principal one has been found the strength of the external tidal 
forces exerted by the neighbor halos, which tend to interrupt the infall of satellites and the accretion of matter particles onto the halo: 
The galactic halos of fixed mass in the stronger tidal fields form on average earlier and in consequence possess less 
substructures \citep{wan-etal11}.

It is not only the tidal strength but also the the configurations of the tidal fields in the surrounding that play a crucial 
role  in the obstruction of the satellite infall and mass accretion into the halos \citep{hah-etal09,shi-etal15,zomg1}.  
Two recent numerical studies based on the high-resolution simulations \citep{zomg1,zomg3} found a significant 
difference in the substructure abundance between the galactic halos located in two different tidal environments. One is inside a 
thick "prominent" filament while the other is at the junction of multiple thin "secondary" filaments. 
For the former case, the satellite infall and matter accretion into a galactic halo is encumbered since the satellites and matter particles 
develop tangential velocities perpendicular to the direction of the elongated axis of the filament, while for the latter case the satellites 
and matter particles can acquire radial motions falling onto the halos \citep{zomg1}. The galactic halos located in the bulky straight 
filaments tend to form early before the complete formation of the filaments, which leads to smaller number of substructures than 
their counterparts at the present epoch \citep{zomg3}. 

As the main purpose of the previous works that have established the above roadmap was to physically explain 
the assembly bias on the galactic scale as well as a paucity of satellites around the Milky Way \citep{GW07}, 
their focus was primarily put on the galactic halos and their substructures \citep{hah-etal09,mao-etal15}. 
Although it is expected that the formation of the substructures of the cluster halos would be similarly affected by the 
tidal fields, the link between the substructure abundances of the cluster halos and the external tidal field may not be the 
formation epochs unlike for the case of the galactic halos.  
It was indeed numerically found that the formation epochs of the cluster halos were nearly independent of the strength of 
the external tidal forces and that the formation epochs (or equivalently the concentration parameters) of cluster halos at fixed 
mass did not play the most decisive role in shaping the substructure abundance \citep{gao-etal11}. 
Besides, on the cluster scale it is difficult to make a quantiative distinction between the thick primary and the thin secondary 
filaments unlike on the galactic scale, and thus the results of \citet{zomg1} and \citet{zomg3} cannot be readily extended to the 
cluster scale. 
 
Here, we take a different approach to address the issue of the tidal effect on the substructure abundance of the cluster halos. 
Introducing a new concept of the {\it orientation coherence} of the surrounding tidal fields and quantitatively distinguishing 
between different tidal environments in terms of the specific sizes of the filaments, 
we will quantitatively investigate how the substructure abundance of the cluster halos depend on the orientation coherence 
of the surrounding tidal fields. 
The organization of this Paper is as follows. Section \ref{sec:filament} presents the procedures to identify the filaments composed 
of the cluster halos from a high-resolution N-body simulation and to determine their specific sizes. Section \ref{sec:substructure} 
presents the measurements of the correlations between the substructure abundance of the cluster halos and the specific sizes of the 
filaments. Section \ref{sec:summary} presents a physical interpretation of the results, discussing the necessary future improvements of 
our analysis. 

\section{Identifying the Filaments of Cluster Halos}\label{sec:filament}

In the previous numerical works which endeavored to quantify the tidal influence on the substructure abundance of dark halos, 
the first step taken was usually to construct the surrounding tidal field either from the dark matter particles (matter tidal field) or 
from the neighbor halos (halo tidal field) \citep[e.g.,][]{hah-etal09,wan-etal11,shi-etal15}. 
In the current work, however, instead of constructing the tidal field,  our first step is to identify the filaments composed of cluster halos 
by employing the adapted minimal spanning three (MST) algorithm \citep{col07,PL09a,PL09b,SL13,alp-etal14}. 
An advantage of our approach is that it can reduce the risk of contamination caused by the numerical noise generated during the 
construction of the halo tidal field on the cluster scale. As shown by \citet{wan-etal11}, the substructure abundance of dark halos 
exhibits a stronger correlation with the halo tidal field constructed from the neighbor halos with comparable masses rather than 
with the matter tidal field. 

But, it would suffer considerably from the small-sample statistics to construct the halo tidal field from the spatial distributions of the 
cluster halos due to the rareness of the cluster halos.  Furthermore, during the construction procedure, the halo tidal field is often 
smoothed with a spherical window function, which has an effect of undermining the degree of the anisotropy of the tidal field. Given 
the previous result that the strongest tidal influence on the substructure abundance of dark halos can be found when the surrounding 
tidal field has highest anisotropy \citep{zomg1},  the smoothing process during the construction of the halo tidal field could also obscure 
the results.  Since no smoothing process is involved in the identification of the MST filaments, our approach can avoid this possible 
systematics. Another merit of our approach based on the adapted MST algorithm is its direct applicability to real observations.  
As well explained and compared with the other algorithms in \citet{lib-etal18}, the adapted MST algorithm is capable of automatically 
adjusting the red-shift space distortion effect. Without making any specified assumption of the background cosmology, it is possible 
to extract the filaments out of the spatial distributions of the bound objects observed in redshift space and 
to directly study the tidal influences of the extracted filaments. 

For our analysis, we will make five samples of the cluster halos and their substructures retrieved from the Rockstar Halo Catalog of 
the Big MultiDark-Planck (Big MDPL) simulations \citep{rie-etal13,rockstar,mdpl}. 
The Planck cosmology was adopted for the Big MDPL simulation starting from the initial conditions described by the following key 
parameters: $\Omega_{m}=0.307$, $\Omega_{b}=0.048$, $\Omega_{\Lambda}=0.693$, $n_{s}=0.96$, $\sigma_{8}=0.8228$ and 
$h=0.6777$ \citep{planck14}. 
The comoving linear length of the simulation box ($L_{b}$), the mass resolution ($m_{\rm cdm}$), and the number of the 
CDM particles ($N_{\rm cdm}$) used for the Big MDPL simulations are given as $L_{\rm b}=2.5\, h^{-1}$Gpc, 
$m_{\rm cdm}=2.359\times10^{10}\,h^{-1}\,M_{\odot}$, and $N_{\rm cdm}=3840^3$, respectively.  

Information on the ID, virial mass, spin parameter, comoving position and velocity of the center of mass, parent ID (pID) of each dark 
halo can be retrieved from the Rockstar Halo Catalog of the Big MDPL. If a dark halo has a parent ID equal to $-1$, then it is a distinct 
halo without belonging to a larger halo. Otherwise, it is a substructure contained in a larger host halo and its parent ID represents 
the corresponding ID of its host halo.  
For our analysis, a cluster halo is defined as a distinct Rockstar halo with ${\rm pID}=-1$  
having virial mass of $M_{\rm host}\ge 10^{14}\,h^{-1}M_{\odot}$, while those Rockstar halos with parent ID equal to the ID of a 
given cluster halo are regarded as its substructure.  We put a mass cut of $10^{12}\,h^{-1}M_{\odot}$ on the substructure mass, 
$M_{\rm sub}$, excluding those substructures with masses smaller than this mass-cut, regarding them poorly resolved in the 
simulations \citep{gao-etal11}.

To find the filaments from the cluster halos with the adapted MST technique,  we first apply the friends-of-friends (FOF) group finder 
with a linking length of $20.4\,h^{-1}$Mpc to the spatial distributions of the cluster halos.  The choice of this specific linking length is 
made through investigating the number of the filaments, $N_{\rm fil}$, against the variation of the linking length $l_{p}$ and 
determining the value of $l_{p}$ that maximizes $N_{\rm fil}$. 
Each of the identified FoF groups represents a network system in which the member cluster halos are within the realm of 
gravitational influence from one another.  Starting from the position of a member halo located in the group center \citep{alp-etal14}, 
we find its nearest member halo, connect the two halos by a straight line, and repeat the same process till all of the member 
halos are connected to construct a MST for each FoF group. Then, through pruning each MST at the level of $p=3$, 
we identify a filamentary connection that stands out with the most linear pattern. 
A detailed description of the adapted MST algorithm and its pruning process, see \citet{SL13} and \citet{alp-etal14}. 

Each identified filamentary network of the cluster halos, which we will call simply a filament from here on, corresponds to a site
where the tidal field peaks both in its strength and its degree of anisotropy on the cluster scale. Suppose that a cluster halo is found to 
reside in a filament. The strongest tidal influence that this cluster halo receives is from the other member cluster halos belonging 
to the same filament.  Hence, the substructure abundance of this cluster halo is naturally expected to depend on the physical properties 
of the host filament such as its richness, spatial extent, shape and etc. 

In Section \ref{sec:substructure}, we will statistically examine how the substructure abundance of a member cluster halo depends on 
the {\it specific size} of its host filament, $\tilde{S}$, which is defined as the spatial extent of a filament divided by its node number, 
where a node number refers to the number of its member cluster halos. A filament forms in a region where the tidal field has two 
positive and one negative eigenvalues. The gravitational stretch of matter distribution will occur along the direction of the 
eigenvector corresponding to the negative eigenvalue (the minor principal axis) while the other two (major and intermediate) 
principal axes coincide with the directions for the gravitational collapse \citep[see][and references therein]{for-etal09}. 
A large-scale coherence of the tidal fields in the orientations of their minor principal axes will lead to the formation of a spatially 
extended rich filament.  We will adopt the specific sizes of the filaments as a practical measure of this {\it orientation coherence} 
of the tidal field. We will adopt the following formula given by \citet{SL13} to determine the specific size, $\tilde{S}$, of each filament. 
\begin{equation}
\tilde{S} = \frac{1}{N_{\rm node}}\left[\left(\Delta x\right)^{2} + \left(\Delta y\right)^{2} + \left(\Delta z\right)^{2}\right]^{1/2}\, ,
\end{equation}
where $\Delta x$, $\Delta y$ and $\Delta z$ represent the ranges of the three Cartesian coordinates of the member cluster halos.

\begin{figure}
\begin{center}
\includegraphics[scale=0.45]{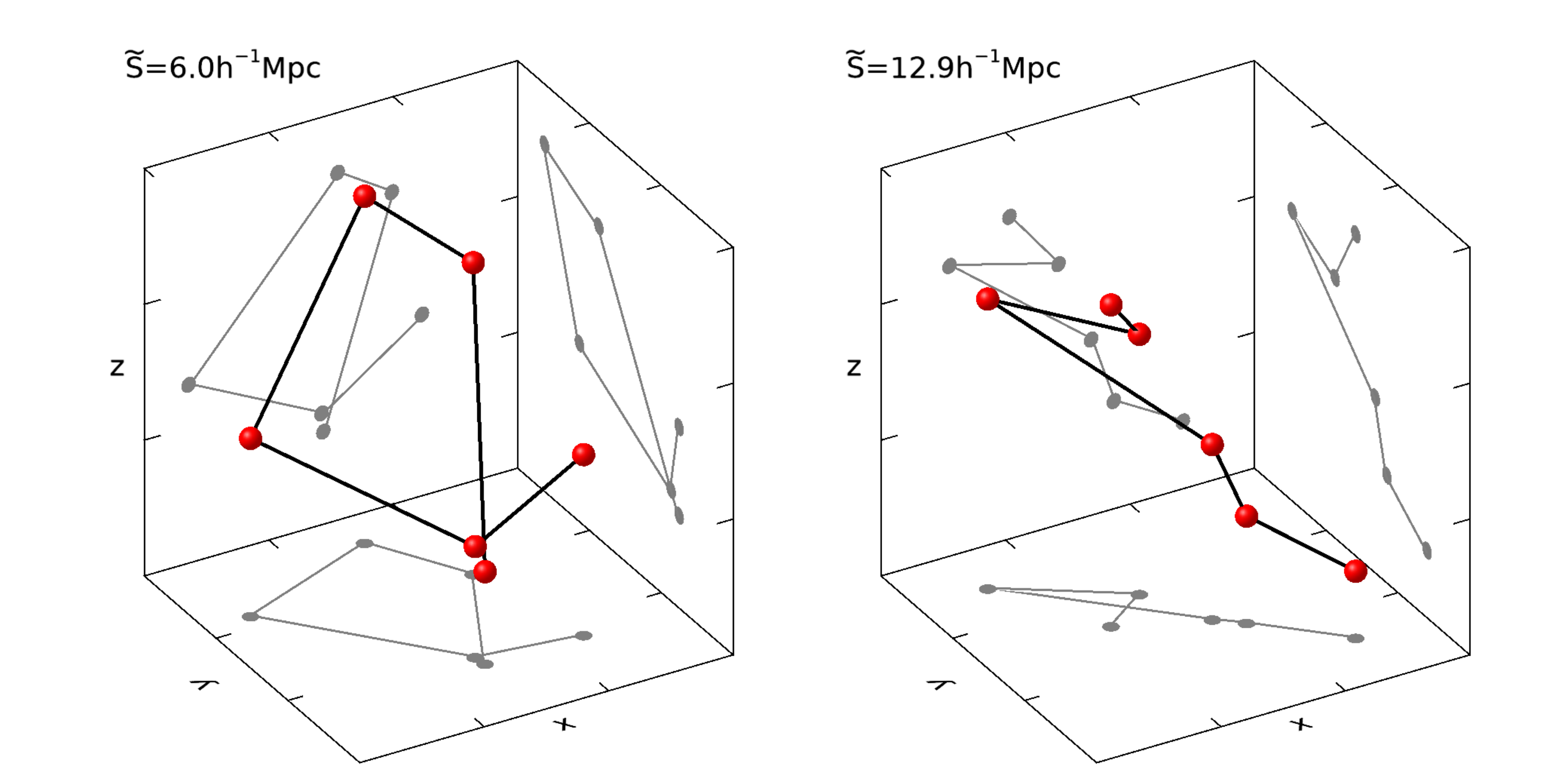}
\caption{Configurations of the cluster halos in the filaments for two different cases of the specific sizes.}
\label{fig:illust}
\end{center}
\end{figure}

\begin{figure}
\begin{center}
\includegraphics[scale=0.65]{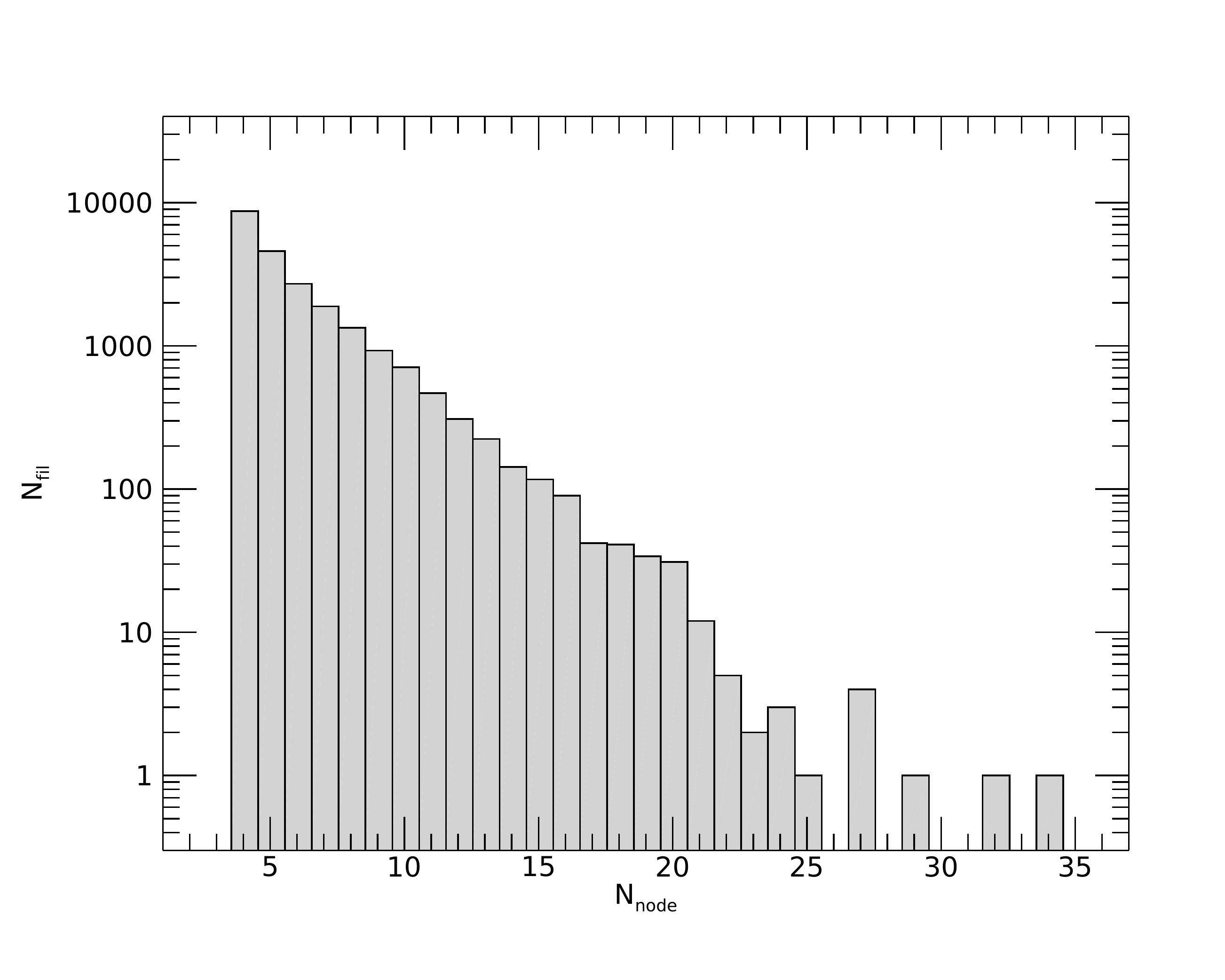}
\caption{Number counts of the filaments as a function of the node numbers.}
\label{fig:nfil_node}
\end{center}
\end{figure}

Figure \ref{fig:illust} compares two identified filaments which differ from each other in their specific sizes. 
As can be seen,  the filament with larger specific size is more elongated and straight than the other filament with smaller specific sizes.  
Figure \ref{fig:nfil_node} plots the number counts of the filaments, $N_{\rm fil}$, as a function of their node numbers, $N_{\rm  node}$, 
showing a sharp decrement of the filament abundance with the increment of the node number. 
Those filaments which have less than four nodes are excluded from our analysis since the filaments with $N_{\rm node}\le 3$ 
may have biased distribution of the specific sizes due to the small number of nodes. 

\section{Substructure Abundance of the Cluster Halos in the Filaments}\label{sec:substructure}

\begin{figure}
\begin{center}
\includegraphics[scale=0.65]{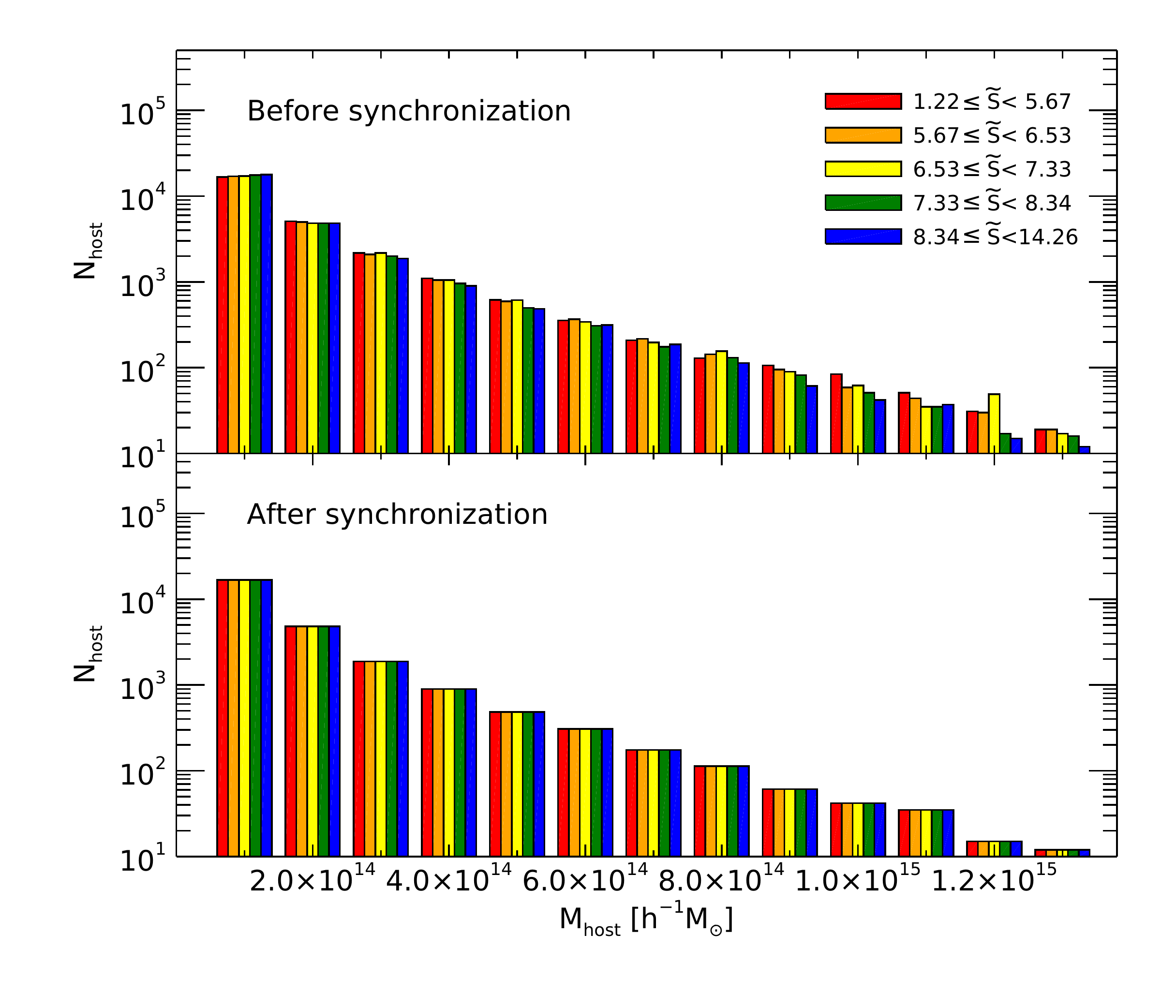}
\caption{Number counts of the cluster halos contained in the filaments as a function of their masses for 
five samples before and after the mass-synchronization process in the top and bottom panels, respectively. 
The specific sizes are in unit of $h^{-1}$Mpc.}
\label{fig:nm_sync}
\end{center}
\end{figure}

Now that the specific sizes of the filaments with $N_{\rm node}\ge 4$ are determined, we classify the cluster halos into five samples by 
the specific sizes of their host filaments and examine the  mass distributions for each sample. The top panel of Figure \ref{fig:nm_sync} 
shows the number counts of the cluster halos belonging to each sample as a function of mass, revealing that the five samples have 
different mass distributions.  To avoid any false signal of $N_{\rm sub}$-$\tilde{S}$ correlation caused by the difference in the mass 
distributions, we deliberately exclude those cluster halos whose masses fall out of the common range to tally the mass distributions of 
the five samples. The bottom panel of Figure \ref{fig:nm_sync} shows the same as the top panel but after the mass-synchronization 
process.  

Let $N_{\rm sub}^{\alpha}$ represent the number of the substructures in the $\alpha$-th cluster halo belonging to a given 
mass-synchronized sample which contains a total of $N_{\rm host}$ cluster halos. Taking the average of the substructure number 
counts over the cluster halos belonging to each mass-synchronized sample 
as $\langle N_{\rm sub}\rangle \equiv \sum_{\alpha} N_{\rm sub}^{\alpha}/N_{\rm host}$, we investigate how the mean number 
of the substructures of the cluster halos changes with the specific sizes of the host filaments. 
The top panel of Figure \ref{fig:nsub_ts} plots $\langle N_{\rm sub}\rangle$ versus $\tilde{S}$ with the Bootstrap errors, from which 
one can witness an obvious trend that $\langle N_{\rm sub}\rangle$ increases as the decrement of $\tilde{S}$. 
Counting only those massive substructures with $M_{\rm sub}/M_{\rm host}\ge 0.1$, we repeat the same analysis, 
the result of which is shown in the bottom panel of Figure \ref{fig:nsub_ts}. 
As can be seen, the massive substructures also retains a signal of the $N_{\rm sub}$-$\tilde{S}$ anti-correlation, albeit  
the errors become larger.

For the results shown in Figure \ref{fig:nsub_ts}, we include all of the identified filaments, regardless of their node numbers. 
However, the node number, $N_{\rm node}$, of a filament can contribute to the $N_{\rm sub}$-$\tilde{S}$ anti-correlation since 
both of $N_{\rm sub}$ and $\tilde{S}$ depend on $N_{\rm node}$.  The specific size of a filament obviously depends on its 
node number by its definition. The abundance of the substructures of a cluster halo is dependent on the node number 
of its host filament since a cluster halo hosted by a filament with higher $N_{\rm node}$ should be exposed to stronger 
tidal effect from more neighbor halos in the same filament. 

To single out the effect of the filamentary tidal influence on the substructure abundance of the cluster halos, 
we now fix the node number of a filament at a certain value and reinvestigate if $N_{\rm sub}$ still retains its anti-correlation 
with $\tilde{S}$. Using those filaments whose node numbers are equal to a fixed value, we reclassify them into five 
subsamples according to their values of $\tilde{S}$ and resynchronize their mass distributions in a similar manner, and 
calculate $\langle N_{\rm sub}\rangle$ for each subsample. Figures \ref{fig:nsub_ts_nd4}-\ref{fig:nsub_ts_nd5} display 
the results for the cases of $N_{\rm node}=4$ and $5$, respectively.  Noting, that for the cases of the higher node numbers, 
the results suffer from large errors due to the small number of the filaments with more than five nodes, we consider 
only the two case of $N_{\rm node}=4$ and $5$.
As can be seen, even when the node number is fixed, the mean substructure 
abundance of the cluster halos on average exhibits a clear signal of anti-correlation with the specific sizes of the host filaments. 
Although the anti-correlation signal becomes weak at fixed node number for the massive substructure case 
of $M_{\rm sub}/M_{\rm host}\ge 0.1$, the results shown in Figures \ref{fig:nsub_ts_nd4}-\ref{fig:nsub_ts_nd5} reveal that 
the dependence of $N_{\rm sub}$ on the number of the cluster halos in the filaments is not the main contribution to the observed 
anti-correlation signal.

\begin{figure}
\begin{center}
\includegraphics[scale=0.65]{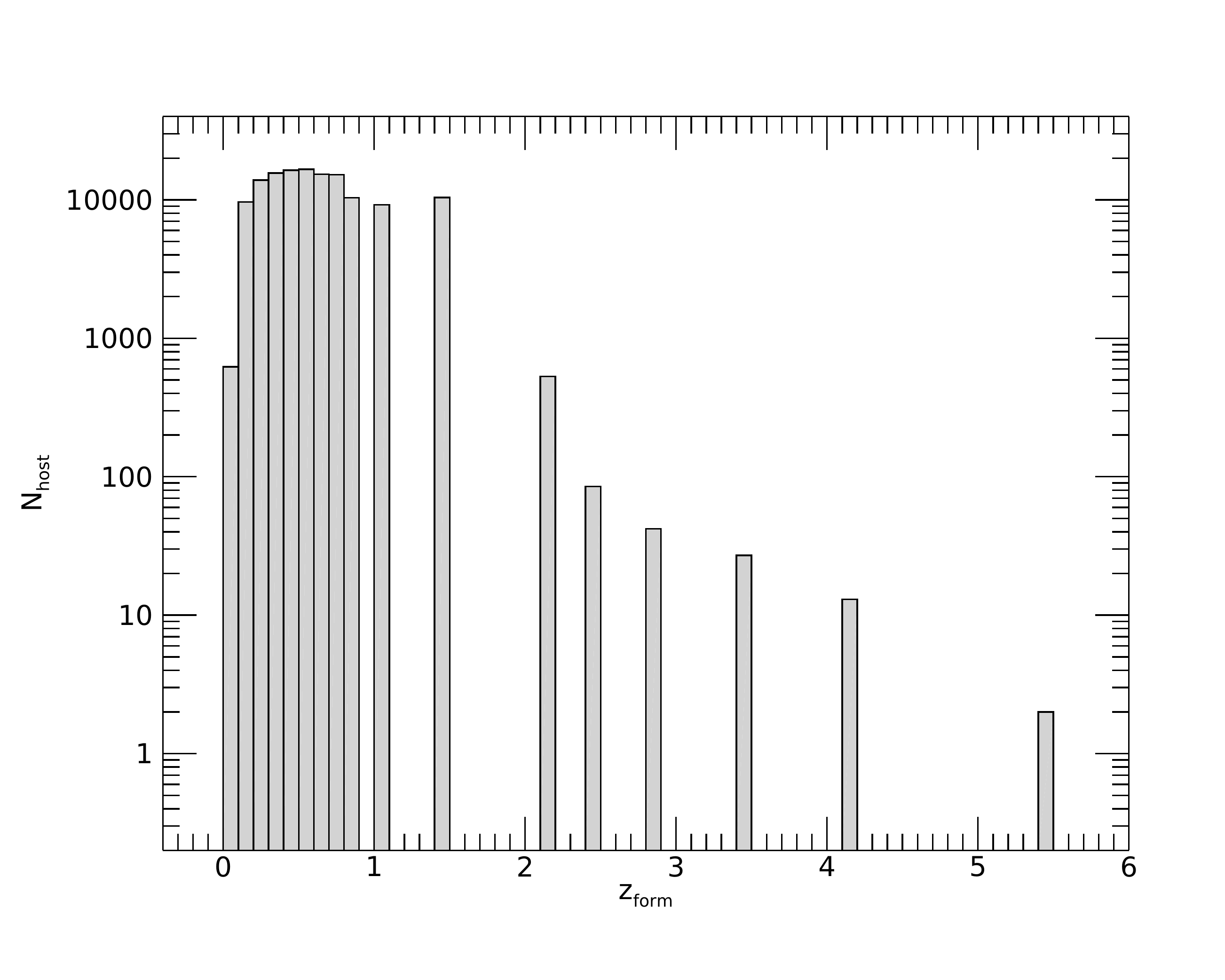}
\caption{Number distribution of the cluster halos as a function of their formation epochs.}
\label{fig:nhost_zform}
\end{center}
\end{figure}

Now that the substructure abundance of the cluster halos in the filaments is found to be anti-correlated with 
the specific sizes of the filaments, we would like to understand what originates this anti-correlation. The usual suspect is 
the assembly bias, i.e., the difference in the formation epochs of the cluster halos as for the case of the galactic halos \citep{zomg3}. 
If at fixed mass scale the cluster halos in the filaments with larger $\tilde{S}$ have a tendency to form earlier, it could explain the 
observed $N_{\rm sub}$-$\tilde{S}$ anti-correlation since those halos formed earlier are known to have on average less substructures.  
Figure \ref{fig:nhost_zform} plots the number counts of the cluster halos as a function of their formation epochs, $z_{\rm form}$. 
For the majority of the cluster halos, the formation epochs lie in the range of $0.2\le z_{\rm form}\le 0.8$. 

\begin{figure}
\begin{center}
\includegraphics[scale=0.65]{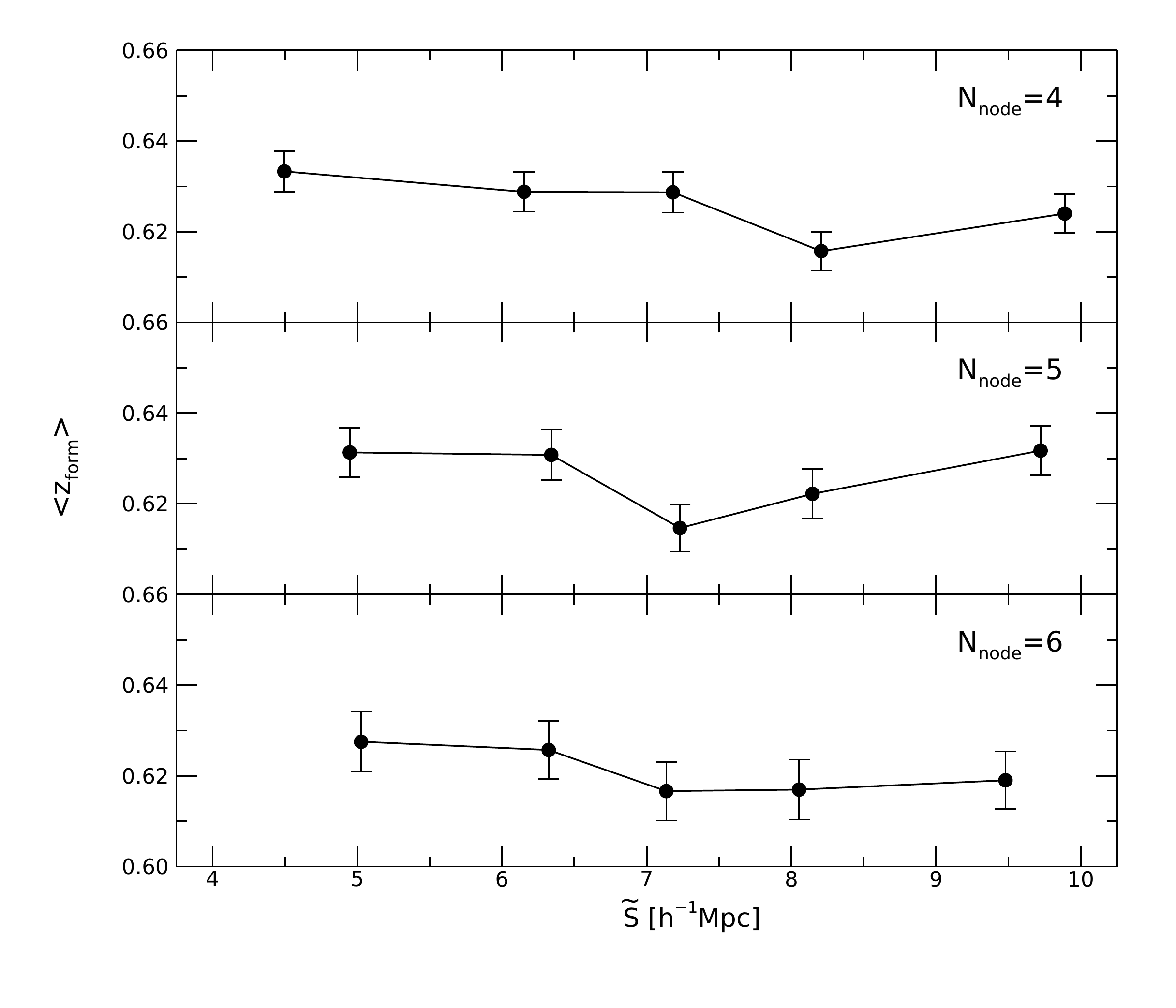}
\caption{Mean formation epochs of the cluster halos vs. the specific sizes of their host filaments.}
\label{fig:zform_ts}
\end{center}
\end{figure}

We compute the mean formation epochs, $\langle z_{\rm form}\rangle$, of the cluster halos belonging to each subsample 
at fixed node number, $N_{\rm node}=4$ and $5$, and display the result in Figure \ref{fig:zform_ts}. 
As can be seen, the mean formation epochs of the clusters seem to depend only quite weakly on the specific sizes of the host 
filaments, which indicates that  the observed signal of the anti-correlation between $N_{\rm sub}$ and $\tilde{S}$ should not be 
produced by the dependence of $N_{\rm sub}$ on $z_{\rm form}$. 
Notwithstanding, to eliminate completely any possible contribution of the $\tilde{S}$-$z_{\rm form}$ correlation 
to the detected signal, we narrow down the range of the formation epochs and reinvestigate the $\tilde{S}$-$N_{\rm sub}$ correlation. 
From each of the five sample, selecting only those cluster halos whose formation epochs fall in a narrow $z_{\rm form}$-interval 
with a width of $\Delta z_{\rm form}=0.1$ to make a subsample  and  synchronizing the mass distributions of the five subsamples, 
we remeasure $\langle N_{\rm sub}\rangle$ as a function of $\tilde{S}$, the results of which are displayed in 
Figures \ref{fig:nsub_ts_zf012}-\ref{fig:nsub_ts_zf089}. As can be seen, narrowing down the range of the formation epochs does not 
erase the anti-correlation signal. 
We also test the survival of the anti-correlation signal against fixing both of $N_{\rm node}$ and $z_{\rm form}$, the result is 
shown Figure \ref{fig:nsub_ts_nd4_zf056} which confirms the robustness of the $N_{\rm sub}$-$\tilde{S}$ anti-correlation. 

Motivated by our detection of the $N_{\rm sub}$-$\tilde{S}$ anti-correlation, we would like to know if and how the 
spatial distributions of the substructures in the cluster halos depend on the specific sizes of the host filaments. Let $r$ be the 
comoving radial distance of a substructure from the center of mass of its host halo whose virial radius is $r_{\rm vir}$. 
Binning the values of the ratios, $r/r_{\rm vir}$, we count the numbers of the substructures at each $r/r_{\rm vir}$ bin, 
$\Delta N_{\rm sub}(r/r_{\rm vir})$, for each cluster halo. Then, we compute the cumulative distribution of the number counts 
of the substructures and then take its average over the cluster halos to determine, $\langle N_{\rm sub}(\le r/r_{\rm vir})\rangle$, 
which is plotted in Figure \ref{fig:cnsub_rrv}. 
The red (blue) solid line corresponds to the case of the cluster halos whose host filaments have the top (bottom) $10\%$ largest 
(smallest) specific sizes. The blurred envelopes of the solid lines represent the corresponding bootstrap errors. 
As can be seen, when all substructures are counted (top panel), the top and the bottom $10\%$ cases make a notable difference 
only at the large radial distance section ($r\ge 0.6r_{\rm vir}$). 
Whereas, when only the massive substructures are counted, the opposite trend is found at the small radial distance section 
($r< 0.6r_{\rm vir}$): the cluster halos belonging to the filaments with bottom $10\%$ smallest specific sizes have more massive 
substructures at shorter radial distances. 

\section{Summary and Discussion}\label{sec:summary}

Given that the substructures of the cluster halos form through the satellite infall and matter accretion and that the majority of the 
cluster halos reside in the the large-scale filaments which are the reservoirs of the satellites and matter particles, 
the formation and evolution of the substructures of the cluster halos are expected to depend on the physical properties of the host 
filaments. We have investigated the correlations between the substructure abundance of the cluster halos and the specific sizes of the 
host filaments, by utilizing the datasets from the Big MDPL \citep{mdpl}.  
The filaments that comprise the cluster halos as nodes have been identified via the adapted MST algorithm \citep{col07} 
and their specific sizes have been determined as their spatial extents per node \citep{SL13}. 
If the tidal fields are coherent over large scales in their orientations of the minor principal axes, then the filaments formed 
through the gravitational stretch of matter along the minor principal axes of the tidal fields are expected to have large specific sizes. 
Hence, the orientation coherence of the tidal field on the cluster scale can be practically quantified by the specific sizes of the 
filaments. 

\begin{figure}
\begin{center}
\includegraphics[scale=0.65]{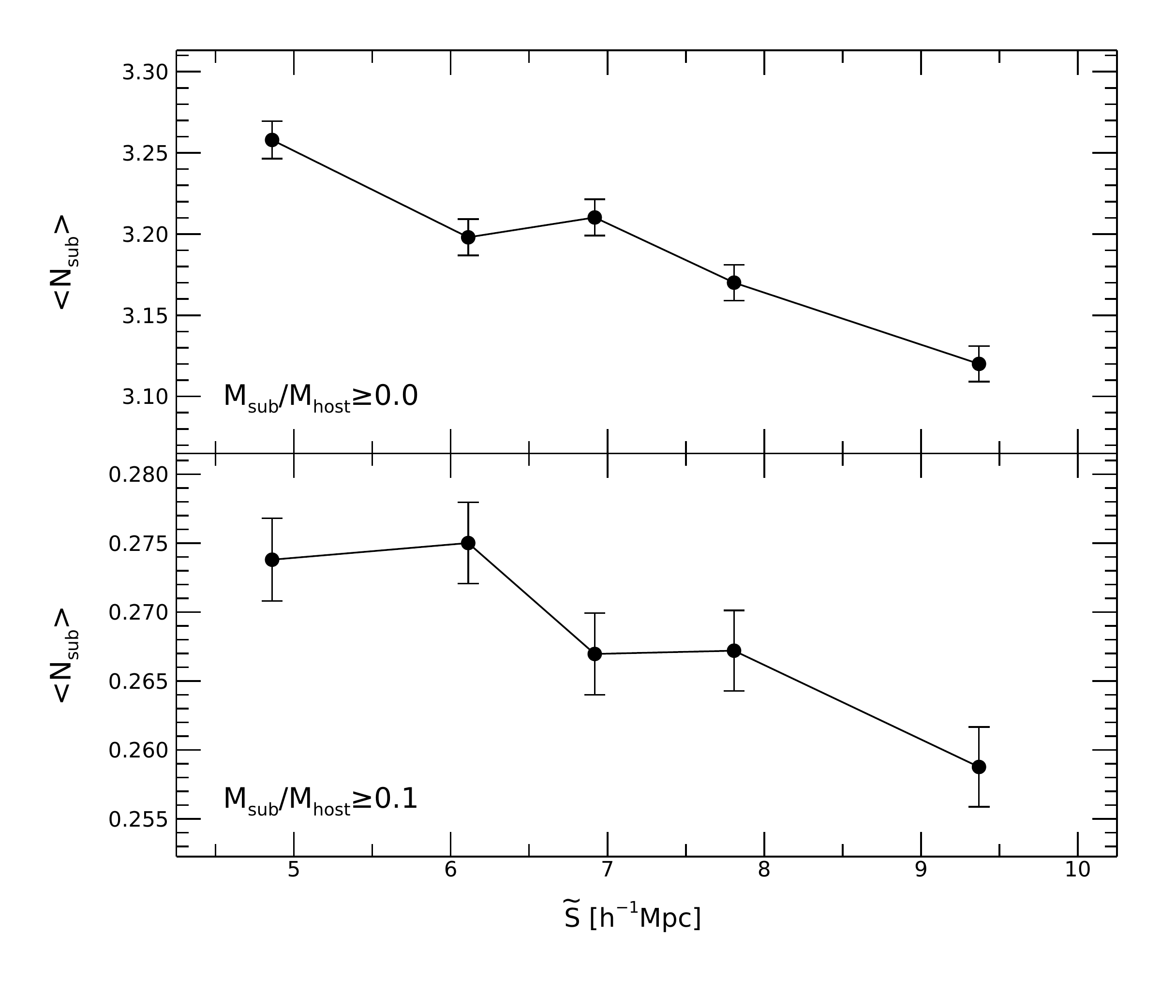}
\caption{Mean substructure abundance of the cluster halos versus he specific sizes of their host filaments.}
\label{fig:nsub_ts}
\end{center}
\end{figure}

\begin{figure}
\begin{center}
\includegraphics[scale=0.65]{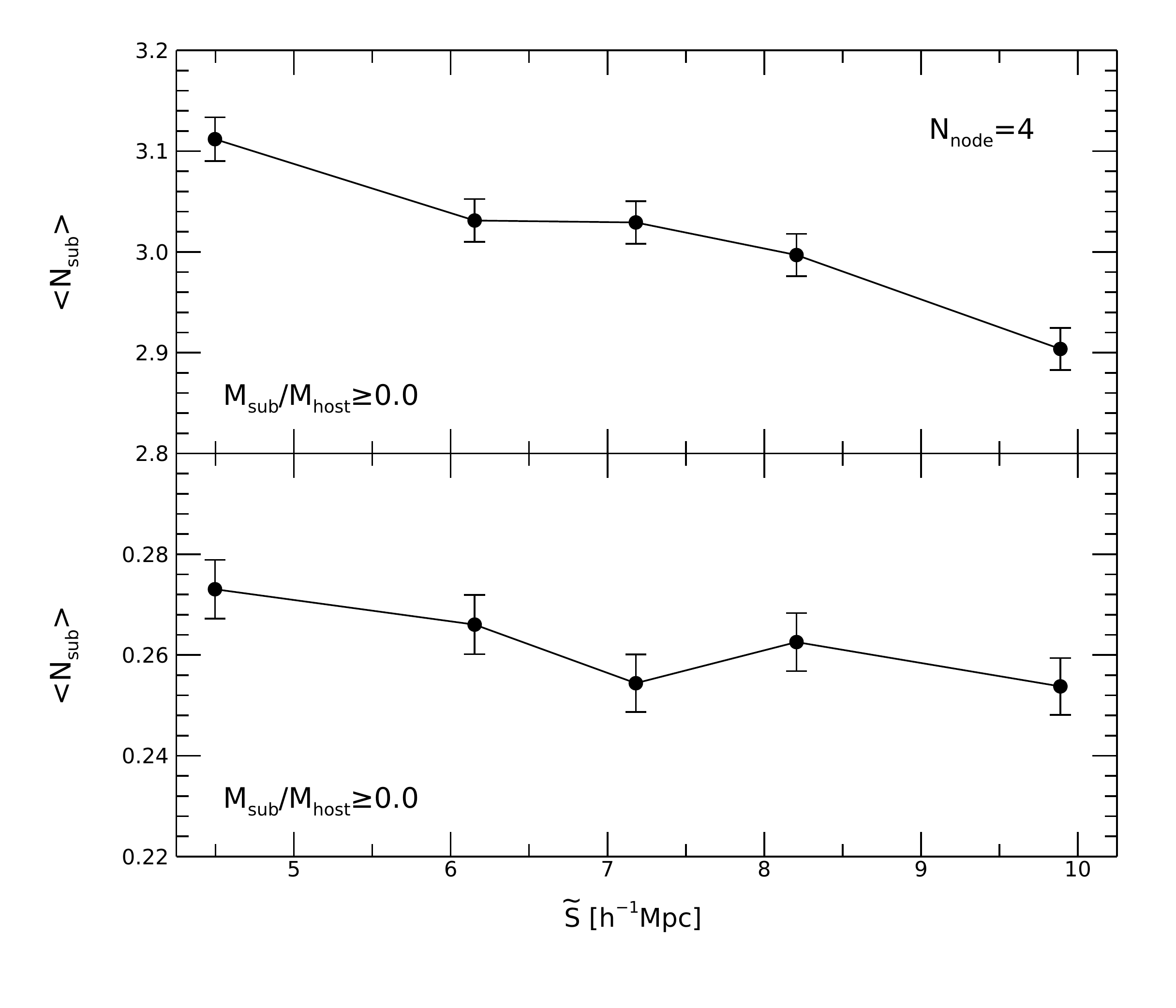}
\caption{Same as Figure \ref{fig:nsub_ts} but for the case that only those clusters whose host 
filaments have a fixed node number of $N_{\rm node}=4$.}
\label{fig:nsub_ts_nd4}
\end{center}
\end{figure}

\begin{figure}
\begin{center}
\includegraphics[scale=0.65]{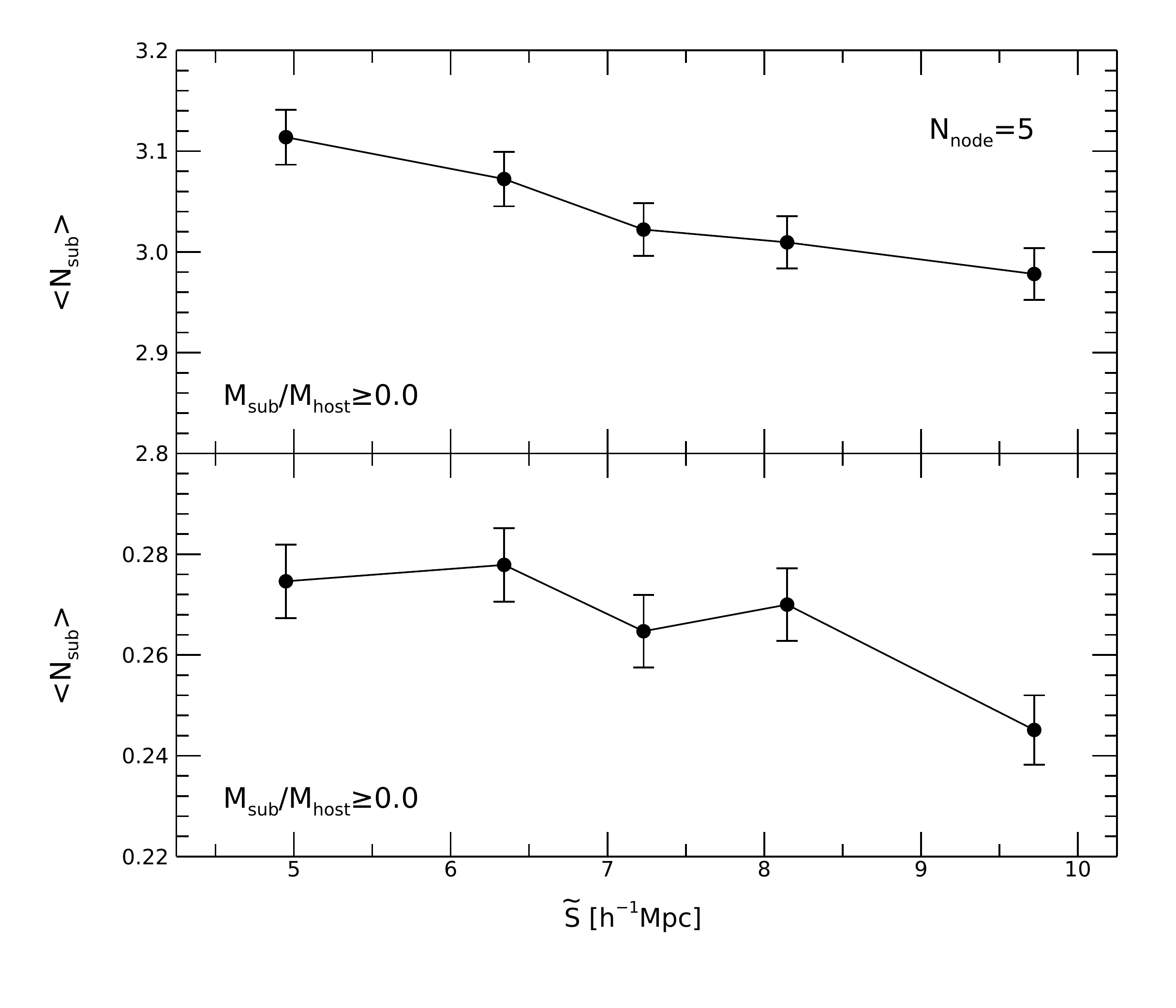}
\caption{Same as Figure \ref{fig:nsub_ts_nd4} but for the case of $N_{\rm node}=5$.}
\label{fig:nsub_ts_nd5}
\end{center}
\end{figure}

\begin{figure}
\begin{center}
\includegraphics[scale=0.65]{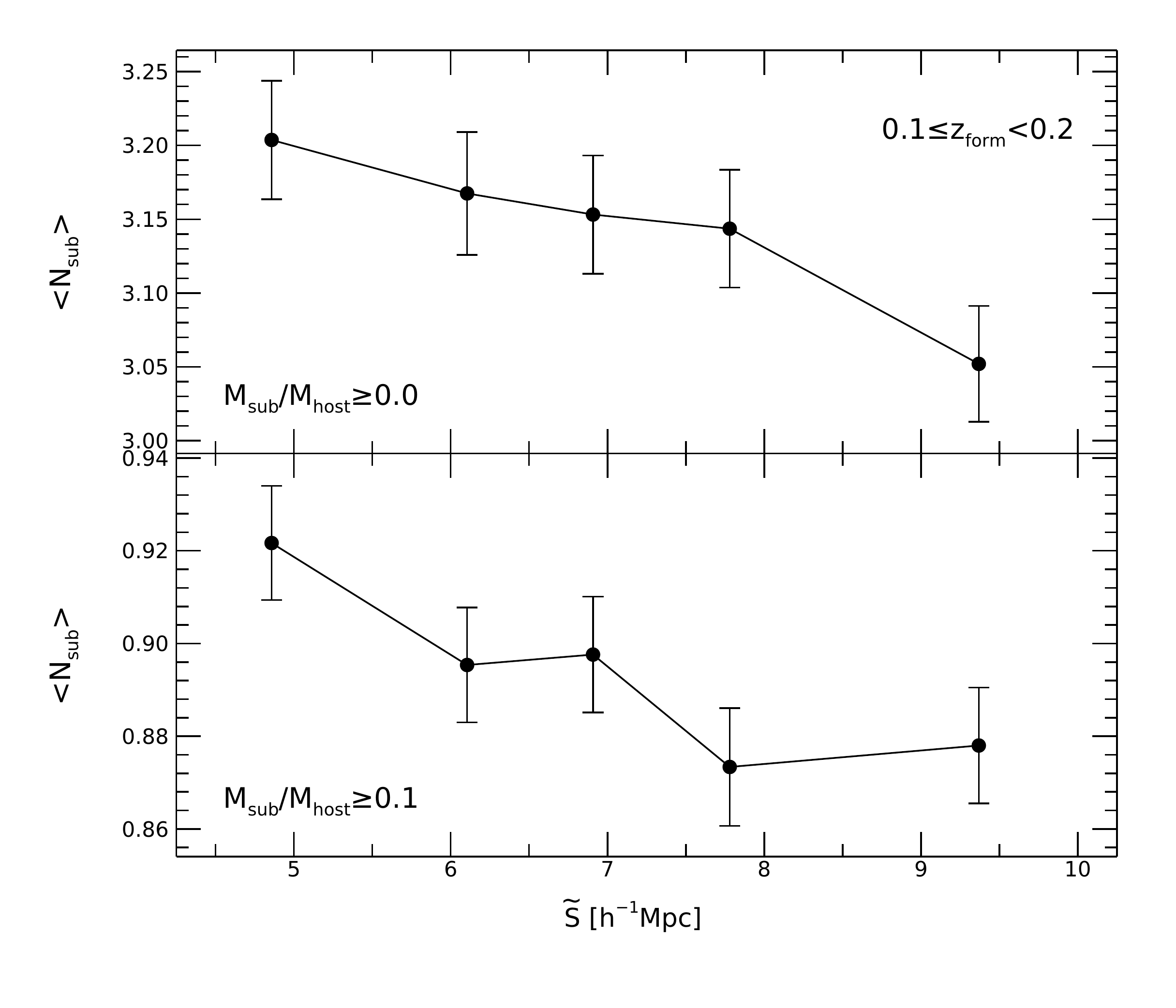}
\caption{Same as Figure \ref{fig:nsub_ts} but for the case of only those clusters with formation epochs 
in the narrow range of $0.1\le z_{\rm form}<0.2$.}
\label{fig:nsub_ts_zf012}
\end{center}
\end{figure}

\begin{figure}
\begin{center}
\includegraphics[scale=0.65]{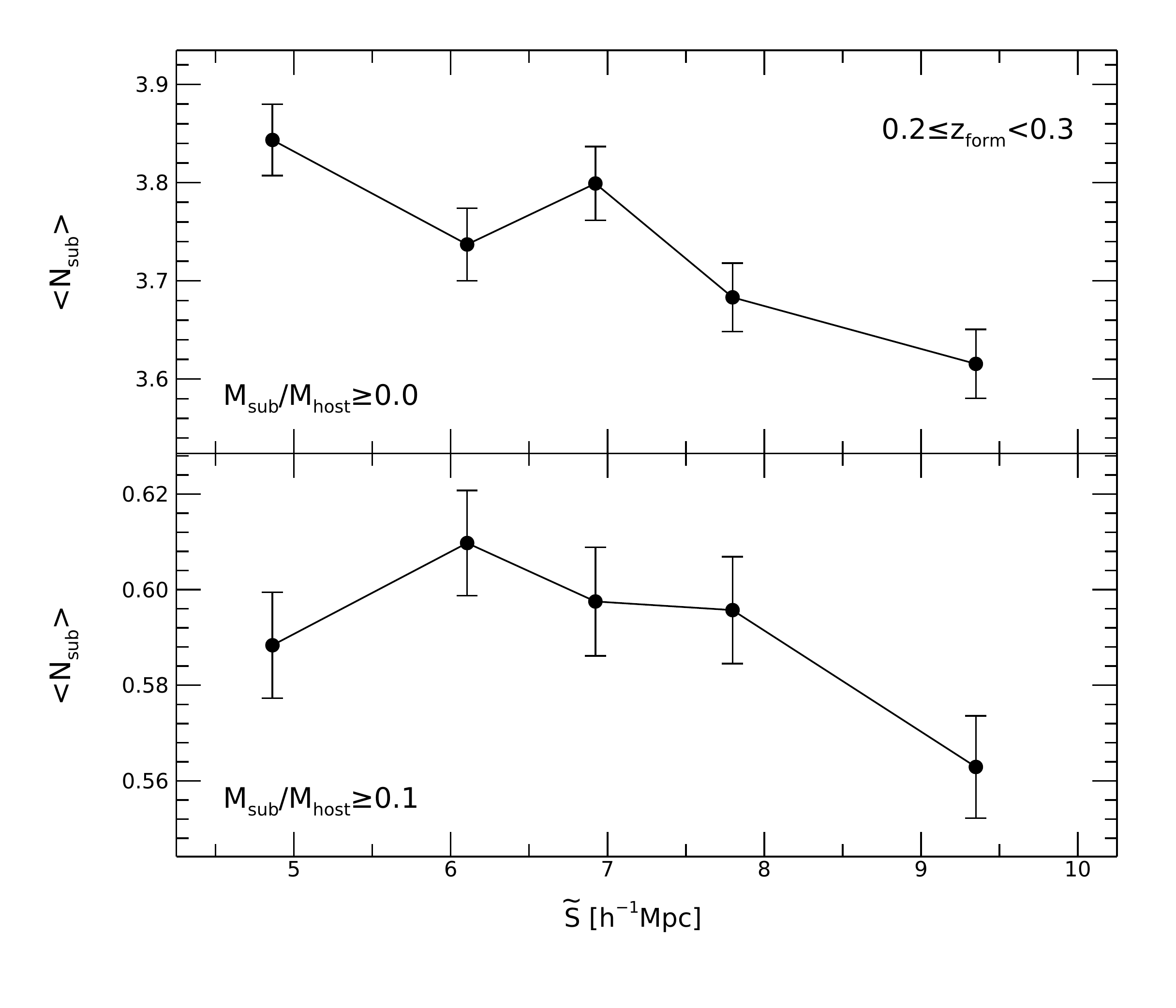}
\caption{Same as Figure \ref{fig:nsub_ts_zf012} but for the case of $0.2\le z_{\rm form} <0.3$.}
\label{fig:nsub_ts_zf023}
\end{center}
\end{figure}

\begin{figure}
\begin{center}
\includegraphics[scale=0.65]{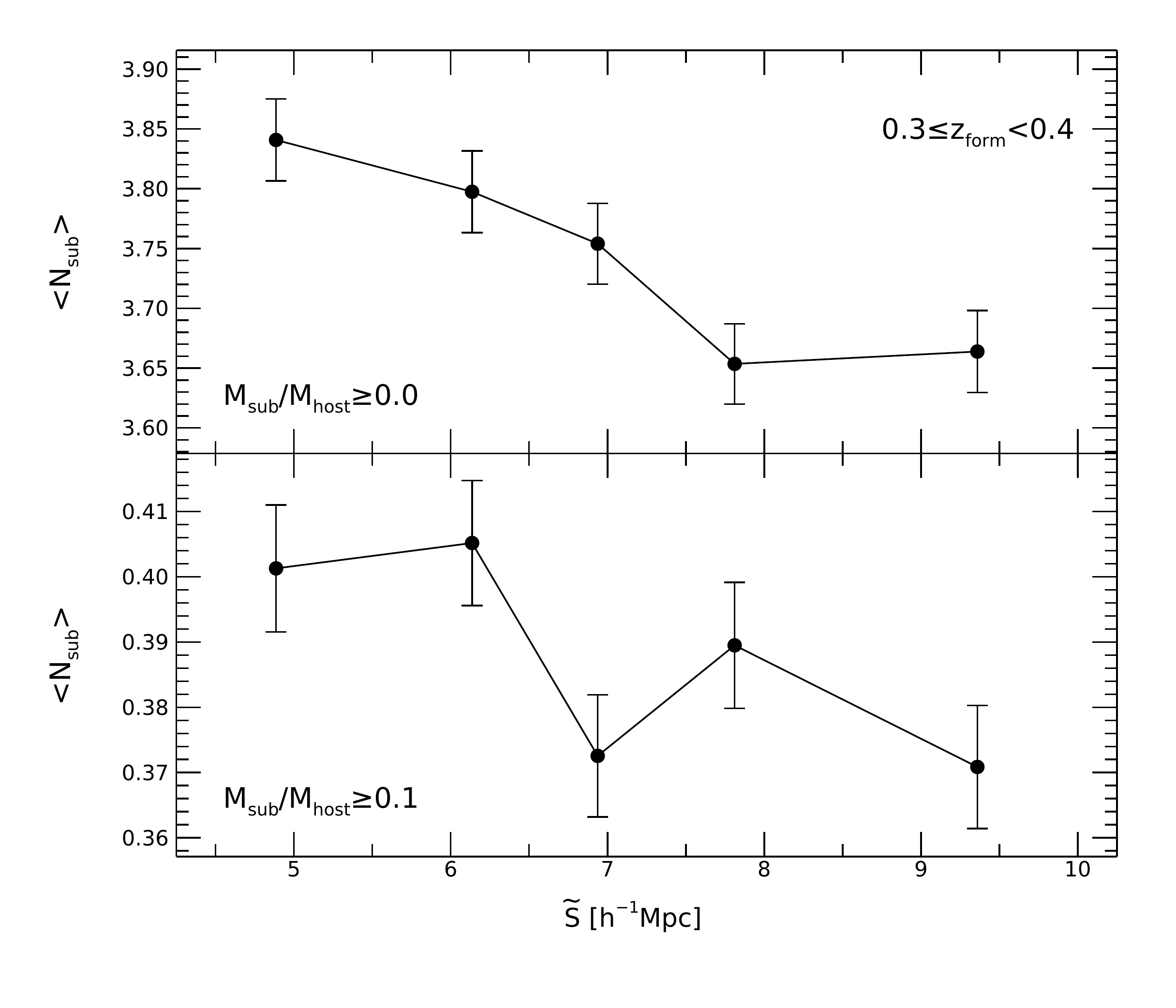}
\caption{Same as Figure \ref{fig:nsub_ts_zf012} but for the case of $0.3\le z_{\rm form} <0.4$.}
\label{fig:nsub_ts_zf034}
\end{center}
\end{figure}

\begin{figure}
\begin{center}
\includegraphics[scale=0.65]{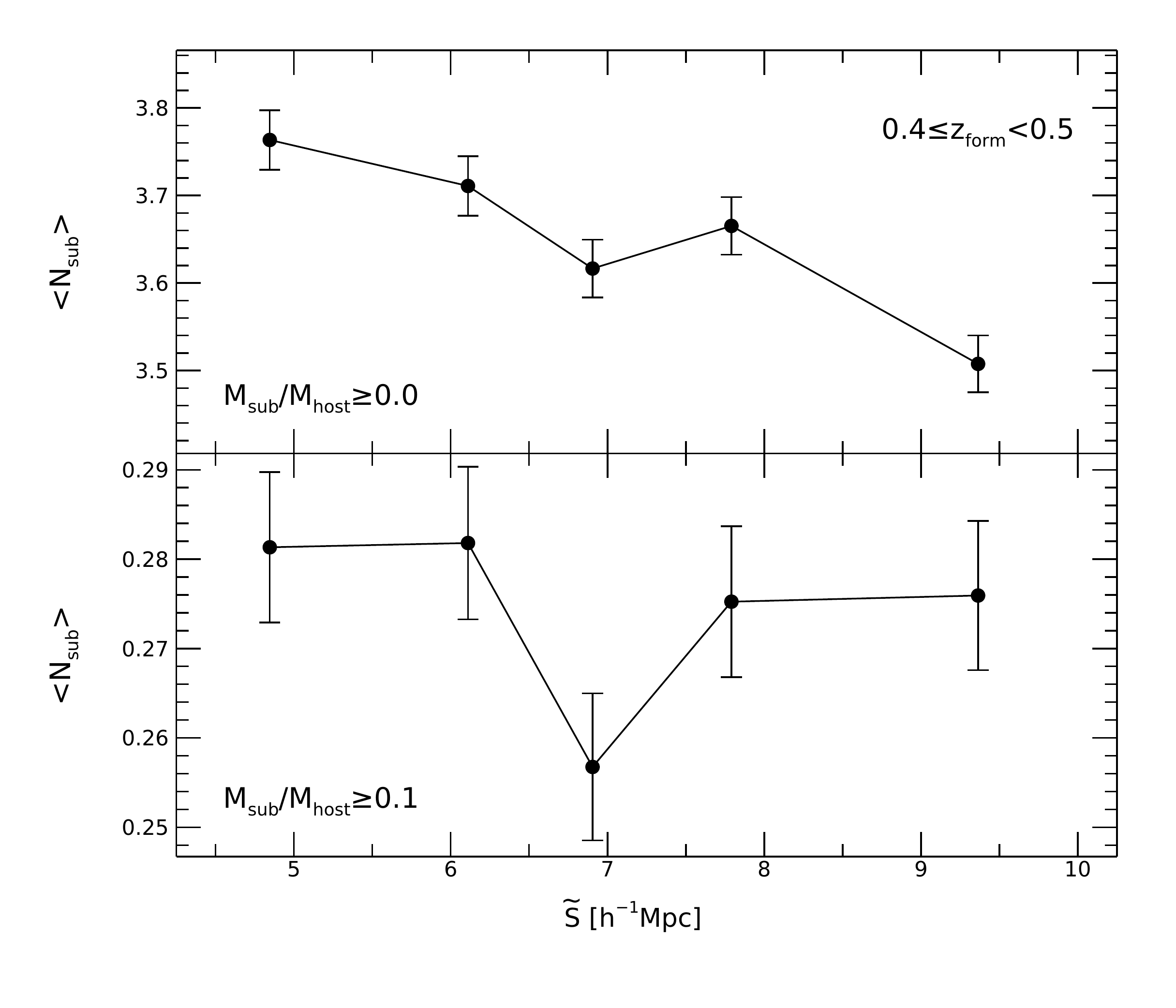}
\caption{Same as Figure \ref{fig:nsub_ts_zf012} but for the case of $0.4\le z_{\rm form} <0.5$.}
\label{fig:nsub_ts_zf045}
\end{center}
\end{figure}

\begin{figure}
\begin{center}
\includegraphics[scale=0.65]{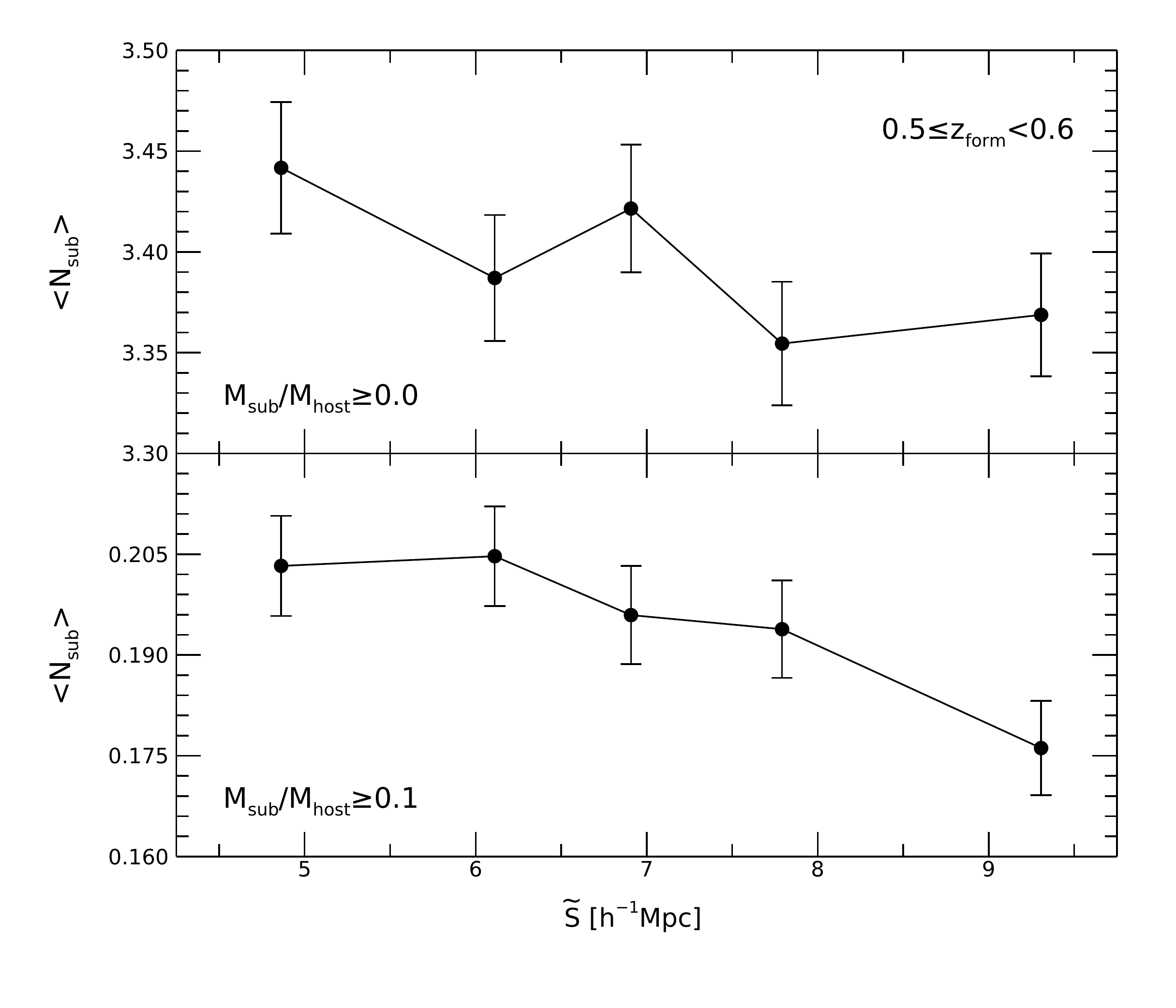}
\caption{Same as Figure \ref{fig:nsub_ts_zf012} but for the case of $0.5\le z_{\rm form} <0.6$.}
\label{fig:nsub_ts_zf056}
\end{center}
\end{figure}

\begin{figure}
\begin{center}
\includegraphics[scale=0.65]{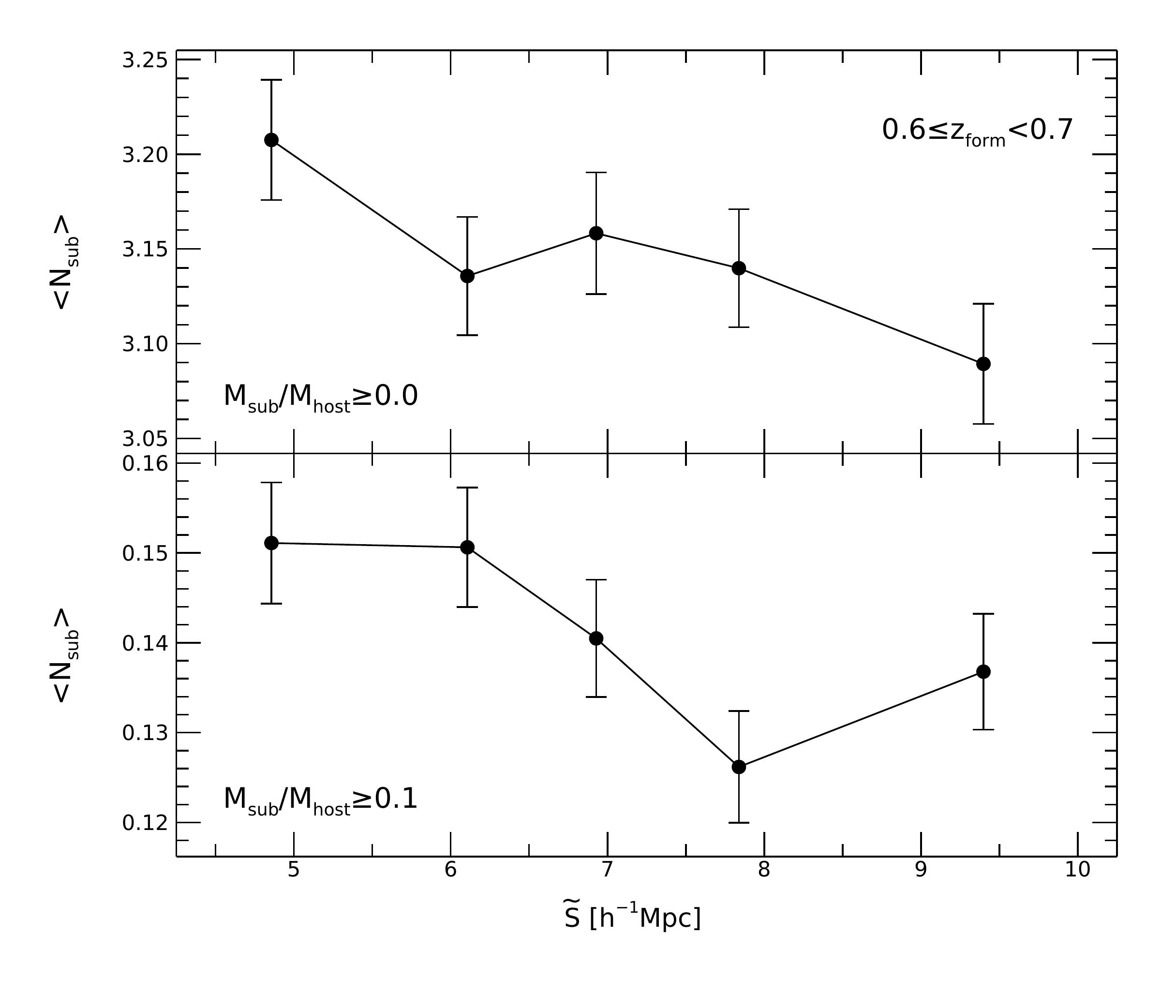}
\caption{Same as Figure \ref{fig:nsub_ts_zf012} but for the case of $0.6\le z_{\rm form} <0.7$.}
\label{fig:nsub_ts_zf067}
\end{center}
\end{figure}

\begin{figure}
\begin{center}
\includegraphics[scale=0.65]{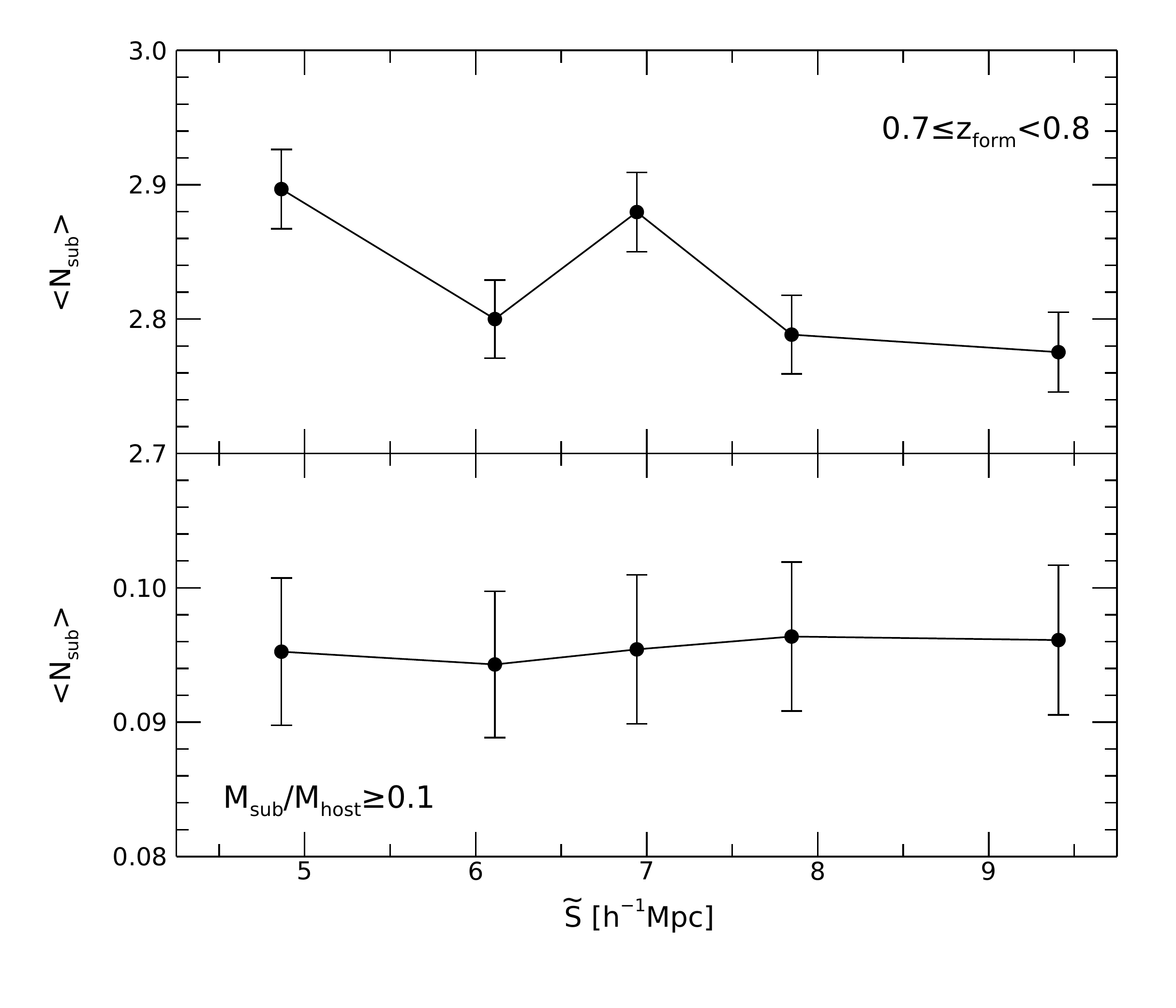}
\caption{Same as Figure \ref{fig:nsub_ts_zf012} but for the case of $0.7\le z_{\rm form} <0.8$.}
\label{fig:nsub_ts_zf078}
\end{center}
\end{figure}

\begin{figure}
\begin{center}
\includegraphics[scale=0.65]{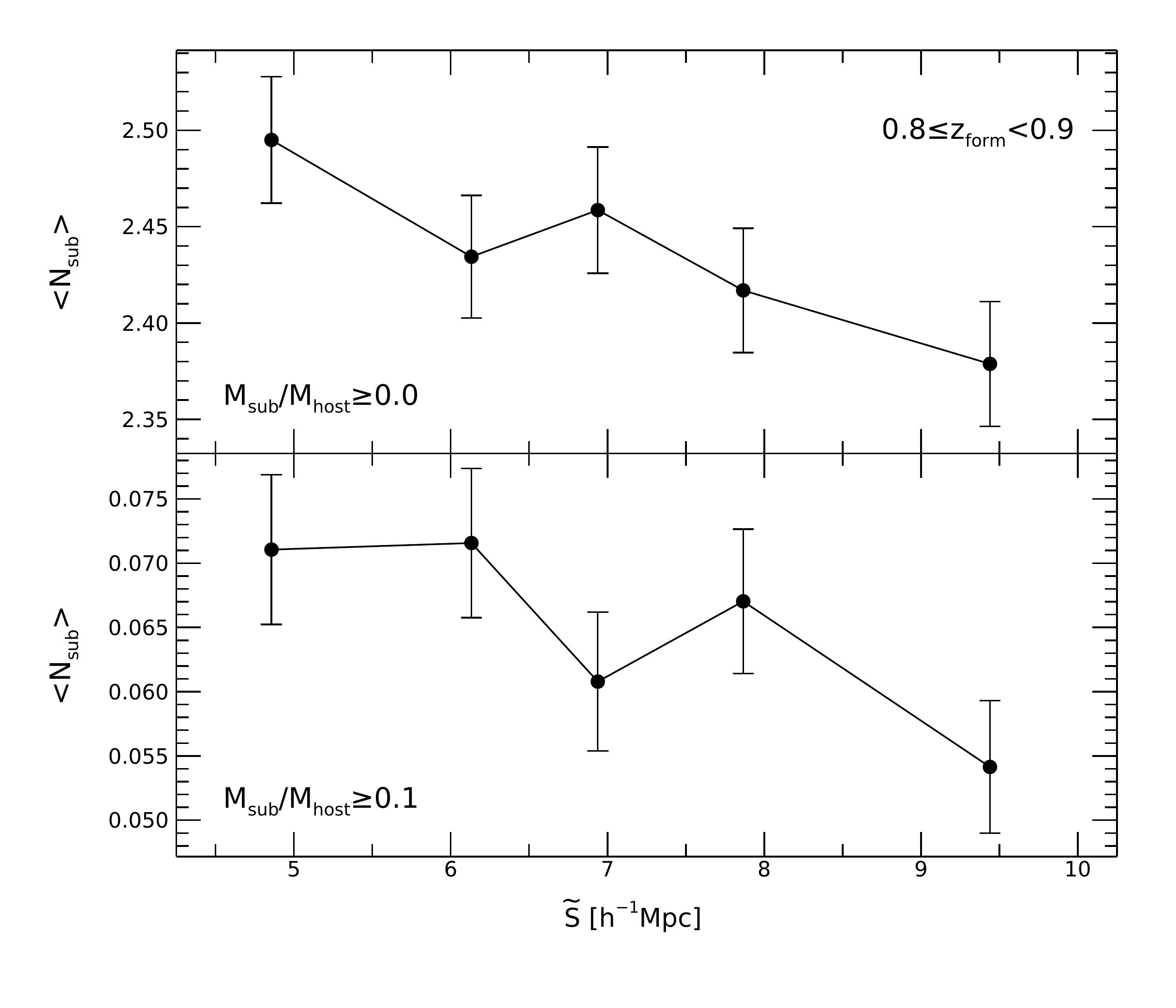}
\caption{Same as Figure \ref{fig:nsub_ts_zf012} but for the case of $0.8\le z_{\rm form} <0.9$.}
\label{fig:nsub_ts_zf089}
\end{center}
\end{figure}

\begin{figure}
\begin{center}
\includegraphics[scale=0.65]{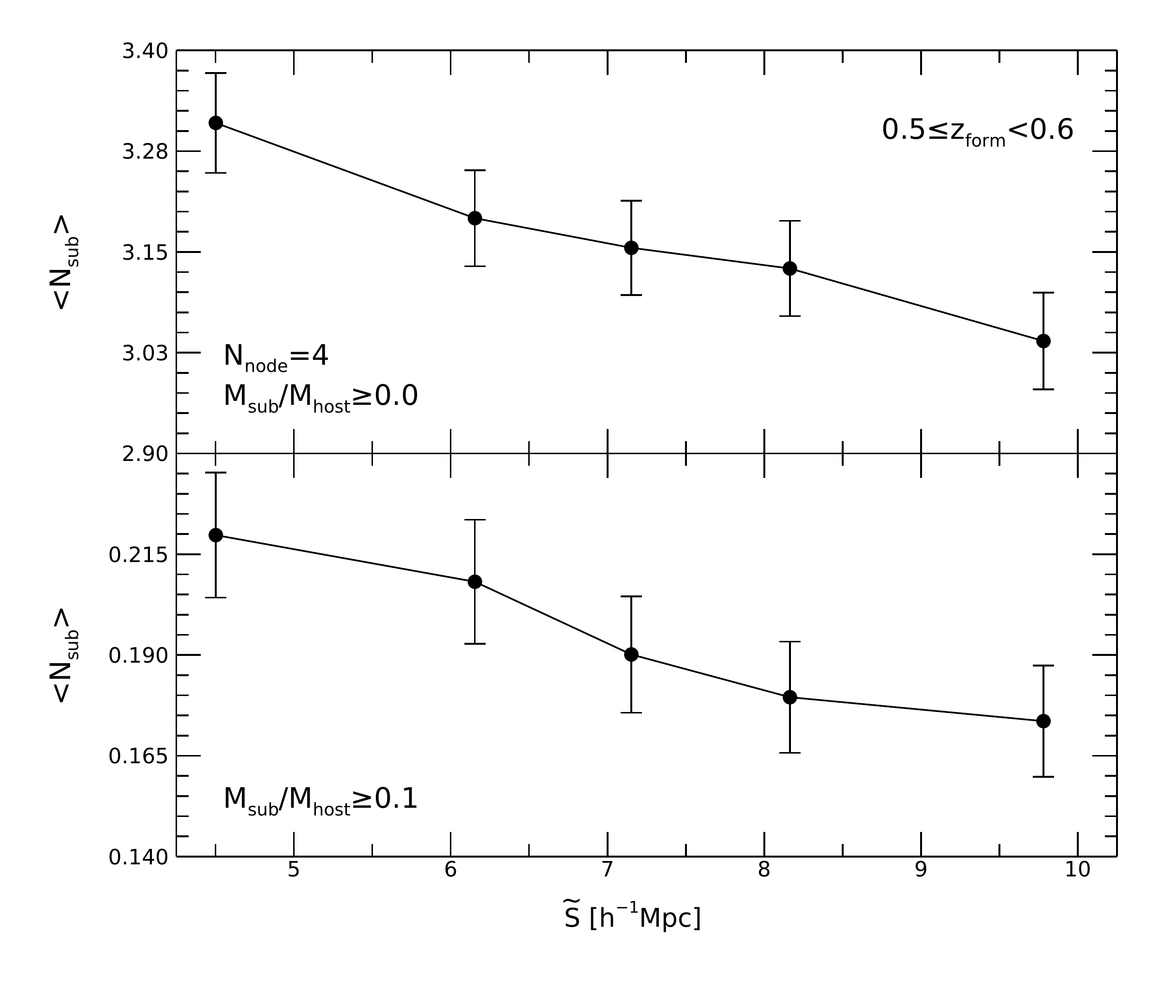}
\caption{Same as Figure \ref{fig:nsub_ts} but for the case of $0.8\le z_{\rm form} <0.9$ 
and $N_{\rm node}=4$.}
\label{fig:nsub_ts_nd4_zf056}
\end{center}
\end{figure}

A signal of anti-correlation between the substructure abundance of the member cluster halos and the specific sizes of their 
host filaments has been detected (see Figure \ref{fig:nsub_ts}). Those cluster halos that constitute the filaments with smaller specific 
sizes have turned out to contain on average more substructures. This anti-correlation signal has been found robust against narrowing 
down the range of the formation epochs of the cluster halos (see Figures \ref{fig:nsub_ts_zf012}-\ref{fig:nsub_ts_nd4_zf056}
as well as against fixing the node numbers of the filaments (see Figures \ref{fig:nsub_ts_nd4}-\ref{fig:nsub_ts_nd5}), 
which implies that the most dominant role in the generation of this anti-correlation should not be the difference in the tidal strength 
among the filaments with different specific sizes

We explain that the anti-correlation should be caused by the obstructed infall of satellites onto the cluster halos in the 
filaments with large specific sizes due to the development of the tangential velocities of the satellites in the confined plane 
perpendicular to the elongated axes of the filaments: When a region evolves to form a filament through its gravitational compression 
along the major and intermediate axes and the gravitational dilation along the minor principal axis of the local tidal field, the satellites 
and matter particles located in the region develop their inward velocities mainly in the plane spanned by the major and the intermediate 
principal axes. In other words, inside a filament, the velocities of satellites and matter particles grow in the directions perpendicular to 
the elongated axis of the filament. Once a filament forms, the satellite infall and matter accretion into the cluster halos that reside in 
the filament should be obstructed since their velocities in the filaments are perpendicular to the direction toward the cluster halos. 
The more coherent the minor principal axes of the tidal fields are in their orientations over large scales, the larger 
specific sizes the filaments would have and thus the more severely the satellite infall and matter accretion should be obstructed. 
For the cluster halos residing in the filaments with small specific sizes, the satellite and matter particles can  
develop radial velocities at non-negligible level in the direction parallel to the elongated axes of the filaments.

\begin{figure}
\begin{center}
\includegraphics[scale=0.65]{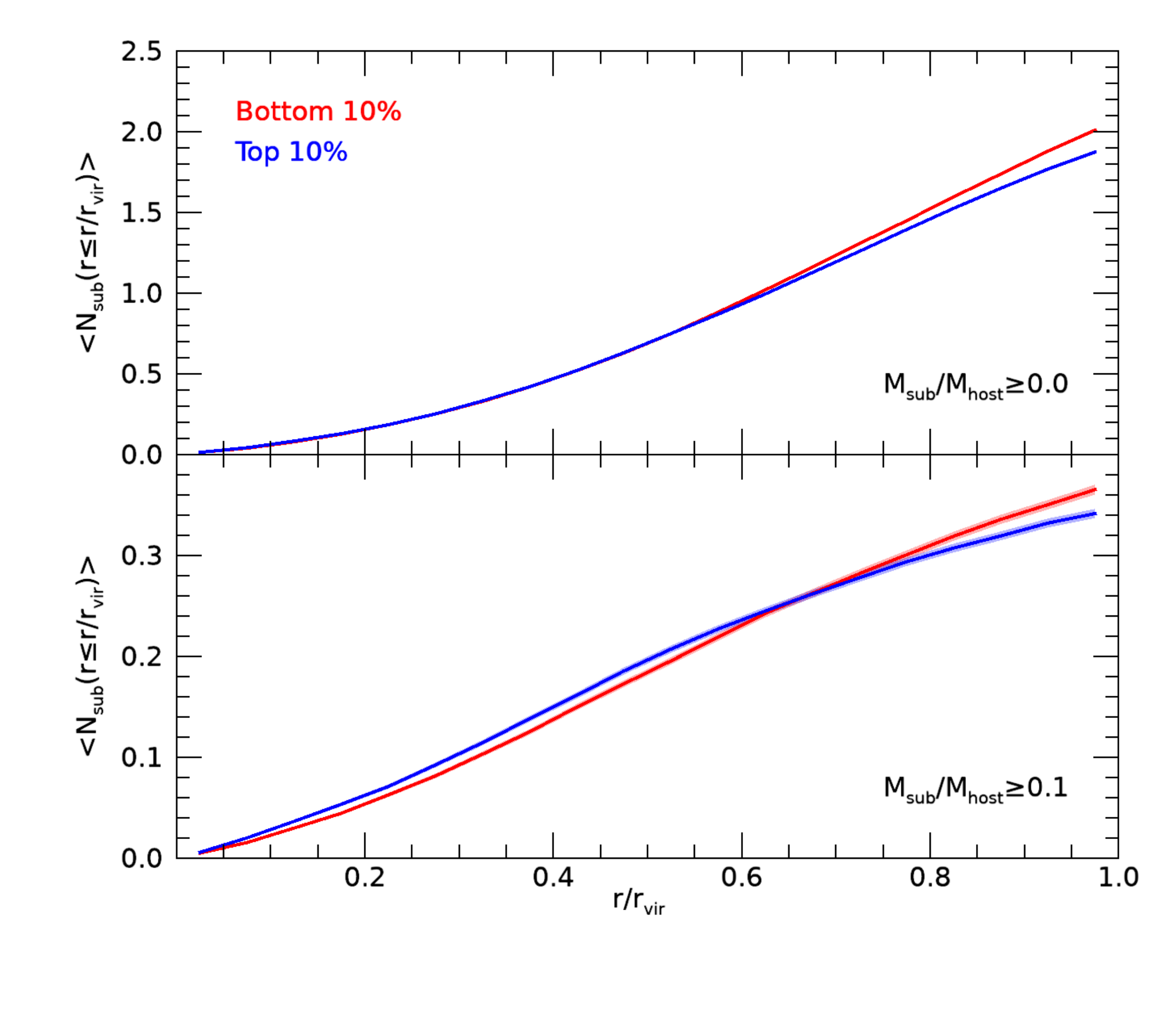}
\caption{Cumulative number counts of the substructures in the cluster halos as a function of their 
rescaled radial distances, $r/r_{\rm vir}$ for the case that the host filaments have the $top 10\%$ 
largest (blue lines) and the bottom $10\%$ smallest specific sizes (red lines), respectively.}
\label{fig:cnsub_rrv}
\end{center}
\end{figure}

We have also determined the cumulative number counts of the substructures, as a function of their radial distances from the cluster 
centers (Figure \ref{fig:cnsub_rrv}). The difference in the substructure abundance between the cluster halos in the filaments with 
large  and small specific sizes has been found to disappear in the small distance section $r\le 0.6 r_{\rm vir}$ 
but to show up in the large distance section $r> 0.6 r_{\rm vir}$. Given that a substructure located at the large 
(small) distances form through relatively recent (early) infall of a satellite,  this result implies that the obstruction of 
the satellite infall and matter accretion becomes effective only after the filamentary connection with high orientation coherence is 
established. In addition,  we have noted that for the case of the cluster halos contained in the filaments with large (small) 
specific sizes the massive substructures with $M_{\rm sub}\ge 0.1M_{\rm host}$ are placed in the small (large) distance section. 
We speculate that the origin of this phenomenon is the difference between the tidal stripping effect of the cluster halos embedded 
in the filaments with small and large specific sizes. In the former (latter) the massive substructures are more (less) severely stripped 
off in the inner region of their hosts. 

Our result is consistent with the finding of \citet{zomg3} that the galactic halos belonging to "prominent" filaments possess on average 
less substructures than those located at the junction of multiple secondary filaments. They used the thickness as a criterion for the 
distinction of a prominent filament from a secondary one. According to their qualitative description, if a filament is straight and thicker 
than the size of a galactic halo,  it is the prominent one where the development of the tangential velocities obstruct the satellite infall 
and matter accretion onto the member galactic halos.  On the cluster scale, however, most of the filaments are as thick as the halo size. 
Therefore, the thickness cannot be used to determine whether a given filament is prominent or secondary. 
In the current work dealing with the cluster halos, we have make a quantitative distinction between prominent and secondary filaments 
in terms of their specific sizes and detected a robust signal of the anti-correlation between the substructure abundance  of the cluster 
halos and the specific sizes of the host filaments, which implies that the orientation coherence of the surrounding tidal field reflected by 
the specific sizes of the filaments plays a key role in regulating the growth of the cluster halos by obstructing the satellite infall and 
matter accretion. 

Yet, to prove the claim that the satellite infall and matter accretion onto the cluster halos in the filaments with large specific 
sizes are indeed obstructed by the development of the tangential velocities in the plane perpendicular to the filament axes, 
it will be necessary to investigate how strongly the velocities of the satellites before falling into the cluster halos and the 
elongated axes of the filaments are correlated and how the correlation strength varies with the specific sizes of the 
filaments. In line with this claim, it will be also interesting to figure out what physical properties of the cluster halos 
other than their substructure abundance depend on the specific sizes of the filaments, which will allow us to have a better 
understanding of the effect of the orientation coherence of the large-scale tidal field on the growth of the cluster halos. 
Our future work will be in this direction. 

\acknowledgments

The CosmoSim database used in this paper is a service by the Leibniz-Institute for Astrophysics Potsdam (AIP).
The MultiDark database was developed in cooperation with the Spanish MultiDark Consolider Project CSD2009-00064.
JS acknowledge the National Junior Research Fellowship (2014H1A8A1022479) through the National Research Foundation 
(NRF) funded by the Ministry of Education. JL acknowledge the support by Basic Science Research Program through the NRF 
funded by the Ministry of Education(2016R1D1A1A09918491) and by a research grant from the NRF to the Center for Galaxy 
Evolution Research (No.2017R1A5A1070354).

\clearpage
\clearpage
\clearpage
\clearpage
\clearpage

\clearpage
\clearpage
\clearpage
\clearpage
\clearpage
\clearpage
\clearpage
\clearpage
\clearpage
\clearpage
\clearpage
\clearpage

\end{document}